\documentclass{article}
\usepackage{PRIMEarxiv}
\usepackage{multicol}
\usepackage[utf8]{inputenc} %
\usepackage[T1]{fontenc}    %
\usepackage{hyperref}       %
\usepackage{url}            %
\usepackage{booktabs}       %
\usepackage{amsfonts}       %
\usepackage{nicefrac}       %
\usepackage{microtype}      %
\usepackage{lipsum}
\usepackage{fancyhdr}       %
\usepackage{graphicx}       %
\usepackage{subcaption}
\usepackage{caption}       %
\graphicspath{{figures/}}     %
\usepackage{xr}
\usepackage{placeins}
\usepackage{soul,color}
\usepackage{makecell}

\title{Generative Design of inorganic compounds using deep diffusion language models
\thanks{\textit{\underline{Citation}}: 
\textbf{Dong et. al. Generative design.}} 
}

\author{
 Rongzhi Dong\\
 Department of Computer Science and Engineering\\
  University of South Carolina\\
  Columbia, SC 29201 \\
  \And 
 Nihang Fu\\
 Department of Computer Science and Engineering\\
  University of South Carolina\\
  Columbia, SC 29201 \\
  \And
  Edirisuriya M. D. Siriwardane\\
 Department of Physics\\
  University of Colombo\\
Colombo 00300, Sri Lanka \\
  \And
 Jianjun Hu *\\
 Department of Computer Science and Engineering\\
  University of South Carolina\\
  Columbia, SC 29201 \\
  \texttt{jianjunh@cse.sc.edu} \\
}

\begin{document}
\maketitle

\begin{abstract}

Due to the vast chemical space, discovering materials with a specific function is challenging. Chemical formulas are obligated to conform to a set of exacting criteria such as charge neutrality, balanced electronegativity, synthesizability, and mechanical stability. In response to this formidable task, we introduce a deep learning-based generative model for material composition and structure design by learning and exploiting explicit and implicit chemical knowledge. Our pipeline first uses deep diffusion language models as the generator of compositions and then applies a template-based crystal structure prediction algorithm to predict their corresponding structures, which is then followed by structure relaxation using a universal graph neural network-based potential. The density functional theory (DFT) calculations of the formation energies and energy-above-the-hull analysis are used to validate new structures generated through our pipeline. Based on the DFT calculation results, six new materials, including Ti$_2$HfO$_5$, TaNbP, YMoN$_2$, TaReO$_4$, HfTiO$_2$, and HfMnO$_2$, with formation energy less than zero have been found. Remarkably, among these, four materials, namely Ti$_2$HfO$_5$, TaNbP, YMoN$_2$, and TaReO$_4$, exhibit an e-above-hull energy of less than 0.3 eV. These findings have proved the effectiveness of our approach.

\end{abstract}


\section{Introduction}

Discovering novel synthesizable and stable materials is of fundamental importance to our society. However, chemical innovation is nontrivial. The material composition and structure must satisfy many stringent constraints such as charge neutrality, balanced electronegativity, synthesizability, geometric symmetry, and mechanical stability. Historically, new material discovery relies on expert heuristics and usually is based on the tinkering of existing materials. Several structure generation studies \cite{schmidt2023machine, V2DB} have used brute-force element substitution to generate new structures based on known prototypes. However, the limitation of this permutation-based approach is that it cannot generate new formula prototypes, it can only employ known formulas as templates, facilitating the generation of novel compositions solely through the substitution of elements. With the development of crystal structure prediction algorithms such as CSMPL \cite{CSPML}, TCSP \cite{TCSP}, and ParetoCSP \cite{ParetoCSP}, the generation of chemically stable compositions has emerged as an increasingly critical challenge. Stable compositions play a pivotal role in mitigating the computational demands associated with subsequent stages of analysis.

Other composition generation methods, such as Generative Adversarial Networks (GANs) \cite{dan2020generative} and Transformers \cite{wei2022crystal, fu2023material}, have demonstrated significant potential in the realm of inorganic compound generation that could break the limitations associated with tinkering-based material design. GANs, when trained with stable samples from the Inorganic Crystal Structure Database (ICSD) \cite{ICSD} database, exhibit the ability to produce novel formulas, with an impressive 84.5\% success rate in passing both charge neutrality and electronegativity balance checks, achieved through the co-evolution of both generator and discriminator components \cite{dan2020generative}. On the other hand, Transformers \cite{attention}, using self-attention mechanisms to grasp contextual information, have paved the way for generative models like the Crystal Transformer \cite{wei2022crystal}. Crystal Transformer is a self-supervised, blank-filling language model that specializes in material composition generation. Remarkably, 78.1\% of compositions generated by the Crystal Transformer successfully meet both charge neutrality and electronegativity balance criteria. Building upon this achievement,  Fu et al. \cite{fu2023material} have extended their efforts to employ various transformer-based language models for material composition generation. Their experiments reveal that approximately 91.2\% of compositions generated by the GPT-J \cite{GPT-J} model, trained on the ICSD database, satisfy both charge neutrality and electronegativity balance checks. A salient disparity between GAN-based and transformer-based generation methods lies in their primary focus: GAN-based approaches tend to explore new composition spaces, whereas transformer-based models are inclined towards compositions that closely resemble the training samples.

Different from all the aforementioned generation methods, diffusion-based generative models have recently demonstrated remarkable capabilities in generating images, texts, proteins, molecules, and even videos. The latest advancements in denoising diffusion probabilistic models \cite{ho2020denoising} have elevated their performance to an even higher standard. The denoising diffusion model comprises two principal processes: the forward diffusion process and the reverse denoising process. The forward diffusion process starts from an initial given sample x$_0$, and then adds a small amount of Gaussian noise to this sample during each of the T steps, producing a sequence of noisy samples x$_1$, x$_2$, ..., x$_t$, ...,x$_T$. As the step t increases, the original sample x$_0$ gradually loses its distinguishable features. When T is extremely large, the sample x$_T$ approximates an isotropic Gaussian distribution. Conversely, the reverse denoising process aims to reconstruct a clear sample x$_{T-1}$ from noisy sample x$_T$ by training a network to reduce the noise distributions. By iteratively executing this reconstruction process for T steps, we are supposed to get the original sample x$_0$. To leverage diffusion language models in the domain of material composition generation, it is imperative to convert material formulas into element sequences. Subsequently, our sequence dataset can be employed to train diffusion-based language generation models for the generation of novel formulas, which are then inputted into crystal structure prediction algorithms to further facilitate structure determination. 

Compared to other Transformer-based language generation models \cite{GPT-J, bert, GPT}, the most significant advancement offered by diffusion models lies in their training methodology. The diffusion models are specifically designed to denoise sequences, prioritizing the creation of meaningful content without being constrained by order. In contrast, traditional Transformers are primarily trained for autoregressive language generation, typically adhering to a left-to-right sequence. Furthermore, diffusion models possess the capability to produce sentences of varying lengths, rendering them highly adaptable to diverse tasks especially our formula generation problem, while Transformers frequently produce fixed-length sequences. Unlike Transformers, which heavily rely on syntactic information, diffusion models are trained to generate content without the need for an explicit representation of syntactic relationships between elements. To summarize, the diffusion language model's unique training objective and its capacity to handle variable sentence lengths outperform traditional Transformers.

In addition to the diffusion language model, which is based on denoising diffusion probabilistic models, diffusion models based on noise conditional score networks \cite{NCSN} have found practical applications in diverse domains such as molecular conformation generation \cite{Molecular}, 3D molecular structure generation \cite{Geodiff}, and crystal structure generation \cite{CDVAE}. ConfGF \cite{Molecular} employs random Gaussian noise with varying magnitudes to perturb the interatomic distances of stable molecules, and then trains the noise conditional score network based on the denoising score. After training, this network can used for score estimation for atomic coordinates to guide the conformation generation. The GEODIFF model \cite{Geodiff} treats each atom within a large molecule as an individual particle. These particles progressively diffuse from their initial states to a noisy distribution. Random noises are introduced at each time step to modify the atomic positions. By training a network that simulates the diffusion process, the inverse network could be used to reconstruct the desired geometric distribution from the noisy distribution. In the realm of crystal structure generation, the Crystal Diffusion Variational Autoencoder (CDVAE) \cite{CDVAE} applies the denoising process to generate crystal structures with lower formation energy by adjusting atomic coordinates and updating atom types. It's worth noting that CDVAE generates structures without relying on templates, which are more likely to result in structures with lower symmetry when initiated with random atomic coordinates.

To address the challenges posed by utilizing diffusion models for the direct generation of structures, which often results in low-symmetry structures, as well as to capitalize on the diffusion model's capacity to generate compositions of variable length irrespective of generation order and syntactic constraints, we introduced a diffusion language model-based composition generation, template-based structure prediction, and a potential-based relaxation pipeline. We first represent compositions of inorganic materials as sequences, which can then be used to train two diffusion language models Diffusion-LM \cite{DLM} and Diffusion-BERT \cite{DBERT} to generate new compositions as they are good for modeling sequence patterns. Subsequently, we subject the generated formulas to rigorous scrutiny, employing checks for charge neutrality, electronegativity balance, and oxidation state compliance to ascertain their chemical validity. For those formulas that successfully pass this initial validation, we then use a template-based structure prediction method CSPML \cite{CSPML} to generate hypothetical crystal structures by choosing templates that share the same compositional ratio as the target formula. After structure generation, we employ a potential-based graph neural network M3GNet \cite{M3GNET} to relax these structures based on the atomic information and interatomic forces to make these structures more thermodynamically stable. M3GNet can also calculate the formation energy of each relaxed structure, structures that exhibit negative formation energies are then further verified based on the density functional theory.

\section{Method}

\begin{figure*}[ht] 
    \centering
    \includegraphics[width=0.9\textwidth]{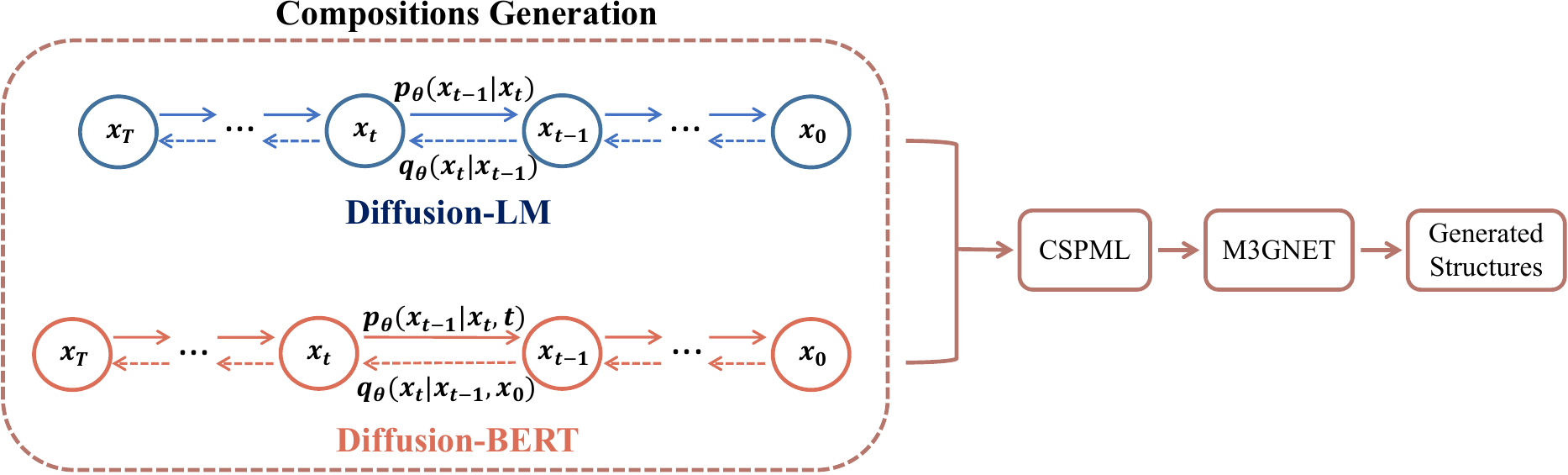}
    \caption{Framework of the diffusion model for generative materials design. Diffusion-LM and Diffusion-BERT are used for composition generation, and the generated compositions are then passed through the CSPML network to get probable structures. These candidate structures are then relaxed by M3GNet and DFT to get the final structures.}
    \label{fig: architecture}
\end{figure*}

Fig \ref{fig: architecture} shows the framework of our diffusion models-based materials generation approach. We employ two different diffusion-based models, namely Diffusion-LM \cite{DLM} and Diffusion-BERT \cite{DBERT}, as composition generators. The generated compositions are then evaluated by checking if they can pass charge neutrality, electronegativity balance, and oxidation state checks. Compositions that successfully pass this evaluation are considered chemically valid. To verify the potential of these chemically valid compositions can be new stable structures, we first use template-based crystal structure prediction algorithm (CSPML) \cite{CSPML} to generate corresponding structures. Subsequently, we utilize a materials graph neural network architecture with 3-body interactions, denoted as M3GNet \cite{M3GNET}, to relax these structures based on a universal potential. Following this relaxation process, we employ density functional theory (DFT) to further validate the stability of the newly discovered materials.

\subsection{Representation of material compositions}

To apply the diffusion language models to our dataset, we need to transform all formulas into sequences, where each atom within the sequence is referred to as a token. Furthermore, we append a period ('.') to the end of these sequences to function as the terminating token for the formulas. For example, we use 'Sr Ti O O O .' to represent SrTiO$_3$. After the exclusion of lanthanides, actinides, and noble gas elements, our dataset comprises 63 distinct elements. Consequently, the vocabulary dictionary necessitates 63 elemental tokens, in addition to six special tokens (which include 'PAD', 'UNK', 'CLS', 'SEP', 'MASK', and '.') traditionally employed in transformer models to represent all sequences \cite{attention}.

\subsection{Diffusion language models for material composition generation}

We trained two different diffusion-based language generation models, Diffusion-LM \cite{DLM} and Diffusion-BERT \cite{DBERT}, from scratch based on our formula sequence dataset and then compared the percentage of chemically valid compositions generated by these two models. These two models are different in both adding noise and denoising processes. In the adding noise process, Diffusion-BERT incorporates the initial state, denoted as x$_0$, in its calculations when determining x$_T$ based on the current state x${T-1}$. Conversely, in the denoising process, Diffusion-BERT computes the next state sample, x$_{t-1}$, by taking into account both the current state x$_t$ and the time step t associated with the current state. In contrast, the Diffusion-LM model derives x$_T$ directly from x$_{T-1}$ by adding Gaussian noises and determines a more clear state x$_{t-1}$ solely based on the current state x$_t$. 

\paragraph{Diffusion-LM}
Most pre-trained large language models are autoregressive models that strictly adhere to the left-to-right order. This constraint becomes especially evident in controllable generation settings such as infilling and syntactic control where global information plays a crucial role. In contrast, Diffusion-LM \cite{DLM} is a non-autoregressive language model that could generate sentences regardless of the order. By iteratively denoises a sequence of Gaussian vectors into word vectors, Diffusion-LM yields a sequence of continuous intermediate latent variables. These continuous latent vectors are then updated by the gradient to make them satisfy the requirements of complex generation tasks. 

\paragraph{Diffusion-BERT}
Diffusion-BERT \cite{DBERT} leverages the capabilities of a pre-trained large language model BERT \cite{bert} as its foundational architecture. This utilization ensures the advantages associated with a well-initialized text generation process. Unlike Diffusion-LM, Diffusion-BERT is a discrete diffusion model. Diffusion-BERT introduced an innovative noise schedule for the forward add noise process, this spindle noise schedule controls the degree of noise added at each step based on the word frequency of each token. The more frequently a word shows up in the training dataset, the earlier it becomes a mask. When denoising from a noise vector, this model tends to generate tokens that most frequently appear in the training stage, which can make the generated sequence more meaningful. 

These two diffusion models represent two different ways of applying diffusion methods within discrete word domains. Diffusion-LM is a continuous diffusion model that transforms the discrete text into a continuous space before adding noise and uses the rounding method after the denoising process to get the discrete test output. In contrast, Diffusion-BERT is trained to directly add noise to each discrete token in the sequence based on the frequency it shows in the training corpus.

\subsection{Crystal structure prediction model for material structure prediction}

After composition generation and duplicate checking, we obtain a material composition candidates dataset. To gain the probable structures of all candidates, we employ a template-based element substitution approach CSPML \cite{CSPML} to select the most similar structure template and then apply element substitution on the selected template structure to generate target structures. CSPML relies on metric learning \cite{Metric} for crystal structure prediction, which can select template structures with high similarity to the given composition from known structure databases. Metric learning uses a binary classifier to distinguish whether two given compositions share similar structures, based on a predefined similarity threshold of local structure order parameters. For a given chemical formula, CSPML initially restricts the candidates to structures with the same compositional ratio (e.g., SrTiO$_3$ has a composition ratio of 1:1:3). Subsequently, the compositional descriptors for both query formula and templates are calculated by XenonPy \cite{XenonPy}, which provides 58 physicochemical features for each element. For any given composition, XenonPy generates a 290-dimensional (58 $\times$ 5) descriptor vector by calculating the weighted mean, weighted sum, weighted variance, min-pooling, and max-pooling of all elements. A traditional multi-layer perceptron is used to figure out the similarity between the template structure's formula and the query formula. The top five template structures with the biggest similarity scores with the query formula are selected as the template candidates. Subsequently, the structure of the query formula is generated by replacing the atoms in the templates with atoms from the query composition. In cases where two or more elements possess the same composition ratio, the choice of substitution element pairs may not be unique. In such instances, the element pairs with the most similar physicochemical properties are chosen. It is important to note that the crystal structure generated through this method shares the same lattice parameters and atomic coordinates as the template structure and requires further relaxation.

\subsection{Graph neural network model for material structure relaxation}

Predicting novel stable crystal structures and their properties with precision remains a crucial objective within the field of computation-guided materials discovery. Although ab initio methods such as DFT have achieved remarkable success in this domain, their considerable computational overhead and limited scalability have constrained their widespread utilization across diverse chemical and structural spaces. To address this challenge, machine learning has been applied to crafting efficient surrogate models capable of predicting material properties on a larger scale. In this paper, after obtaining foundational structures through template-based element substitution techniques, we employ M3GNet \cite{M3GNET}, a potential-based graph neural network method, for structure relaxation. M3GNet combines graph-based deep learning interatomic potential (IAP) with traditional IAPs' many-body features, while also accommodating flexible graph material representations. This model takes position-included graphs as inputs, embedding atomic numbers and pair bond distances as graph features. The graph convolution module then updates bond and atom information based on atom indices and angles calculated through the many-body computation module. Diverging from previous materials graph implementations like MEGNet, M3GNet uniquely incorporates atomic coordinates and the 3 $\times$ 3 lattice matrix in crystals. Remarkably, M3GNet is trained on both stable and unstable structures, enabling it to predict more accurate energy values for unstable materials. The exceptional capacity of the M3GNet-based relaxation algorithm to swiftly and precisely relax diverse crystal structures while simultaneously predicting their energies renders it well-suited for expansive materials discovery endeavors.

\subsection{DFT calculations}

We carried out the first-principles calculations based on the density functional theory (DFT) using the Vienna \textit{ab initio} simulation package (VASP) \cite{Vasp1, Vasp2, Vasp3, Vasp4} to optimize the candidate structures suggested by the machine learning models. The projected augmented wave (PAW) pseudopotentials were used to treat the electron-ion interactions \cite{PAW1, PAW2} with 520 eV plane-wave cutoff energy. The generalized gradient approximation (GGA) based Perdew-Burke-Ernzerhof (PBE) method was considered for the exchange-correlation functions \cite{GGA1, GGA2}. The energy convergence criterion was 10$^{-5}$ eV and the force convergence criterion was 10$^{-2}$ eV/{\AA} for all the DFT calculations. The Brillouin zone integration for the unit cells was performed employing the $\Gamma$-centered  Monkhorst-Pack $k$-meshes. The formation energies (in eV/atom) of the materials were determined employing the formula in  Eq.~\ref{eq: form}, where $E[\mathrm{Material}]$ is the total energy per unit formula of the target structure, $E[\textrm{A}_i]$ is the energy of $i^\mathrm{th}$ element of the material, $x_i$ indicates the number of A$_i$ atoms in a unit formula, and $n$ is the total number of atoms in a unit formula($n=\sum_i x_i$).  The Pymatgen code \cite{Pymatgen} was used to compute the energy above hull values of the materials with negative formation energies.

\begin{equation}
    E_{\mathrm{form}} =\frac{1}{n}(E[\mathrm{Material}] - \sum_i x_i E[\textrm{A}_i])
    \label{eq: form}
\end{equation}

\subsection{Evaluation criteria} 

We employ multiple performance measures to assess the effectiveness of the two diffusion-based material composition generation models. In the evaluation of the diffusion-based material composition generator, we assess its validity, uniqueness, recovery rate, and novelty. For evaluating the template-based structure generator and relaxer, we utilize the formation energy as a key indicator. To further validate these generated structures, we perform energy calculations using VASP to determine their energy above the hull.

\textbf{Validity.} We assess the validity of generated formulas using three key metrics: the percentage of charge-neutral (CN) formulas among all generated ones, computed with the SMACT package \cite{SMACT}; the percentage of formulas with electronegativity balance (EB); and the percentage of formulas passing the oxidation states (OS) check. The OS predictions are facilitated by BERTOS \cite{fu2022composition}, a neural network model comprising a BERT network \cite{devlin2018bert}, and a simple classifier. Trained on extensive material datasets, BERTOS achieves an impressive 97.61\% accuracy in OS prediction, reinforcing its efficacy in assessing the electronegativity balance of input formulas.

\textbf{Uniqueness.} The uniqueness percentage is determined by dividing the number of unique samples by the total generated samples. This metric provides insight into the models' capacity to produce diverse samples.

\textbf{Novelty.} To gauge the models' proficiency in generating novel materials, we compute the novelty of generated formulas. Novelty is quantified by the percentage of generated materials that do not have counterparts in the training samples.

\textbf{Recovery rate.} Another critical evaluation criterion for our models revolves around their ability to generate chemically valid materials that align with known entries in established materials databases. The recovery rate is ascertained by calculating the percentage of known materials successfully rediscovered among a given set of generated samples (10,000 in this study).

\textbf{Formation energy.}  An integral aspect of evaluating the performance of structure generation models involves scrutinizing the stability of generated structures. For structures produced and optimized through our pipeline, we calculate their formation energy using M3GNet.

\textbf{Energy above the convex hull.} The energy convex hull is constructed based on well-established stable structures, as investigated by Liu et al. \cite{liu2015spinel}. Structures with energies residing on the convex hull are deemed thermodynamically stable, whereas those situated above it tend to be either metastable or unstable. Among all structures exhibiting negative formation energies, we utilize the energy above the convex hull as an additional criterion for selecting more stable candidates.

\subsection{Hyperparameters and training}

For formula generation, each diffusion model trained on the MP database generates 10,000 samples. After generation, we use the CSPML method \cite{CSPML} to predict the structures of these composition candidates. M3GNet \cite{M3GNET} is then used to relax the generated structures. Table \ref{table: hyperparameters} shows the hyperparameters used in the Diffusion-LM and Diffusion-BERT models. In both models, we use an element vocabulary of size 69 and the maximum length of generated formula sequences is set to 50. The number of training epochs is set to 20,000. There are some unique parameters used in both diffusion models, including the number of ResNet layers in the Diffusion-LM model and word frequency in the Diffusion-BERT model. To determine appropriate values for these parameters, three ablation studies are established. The results of these ablation studies concerning ResNet layers and word frequency are provided in Table \ref{tab:resnet}, and Table \ref{tab:word_freq}, respectively. We take the percentage of generated compositions passed the charge-neutral (CN) and electronegativity balance (EB) Based on the findings of the ablation studies, the number of ResNet layers in the Diffusion-LM model is set to 6, while the word frequency for the Diffusion-BERT model are set to 0.3, respectively. 

\begin{table*}[ht]
\centering
\caption{Hyperparameters of Diffusion-LM and Diffusion-BERT}
\label{table: hyperparameters}
\begin{tabular}{|c|c|c|c|c|c|c|}
\hline
Model          & Diffusion step & Vocab & Length & ResNet layer &  Word frequency \\ \hline
Diffusion-LM   & 2,000           & 69         & 50              & 6            & N/A           \\ \hline
Diffusion-BERT & 2,048           & 69         & 50              & N/A          & 0.3            \\ \hline
\end{tabular}
\end{table*}

We conducted experiments involving different numbers of ResNet layers within the Diffusion-LM model to assess the proportion of generated samples capable of satisfying the CN+EB check. As illustrated in Table \ref{tab:resnet}, when we progressively increased the number of ResNet layers from 2 to 10, the percentage of generated samples meeting the criteria for chemical stability exhibited an initial rise followed by a subsequent decline. Notably, we identified that the model attains its peak performance, amounting to 43.84\% when employing 6 ResNet layers.

\begin{table}[ht]
\centering
\caption{Ablation studies for ResNet layers used in Diffusion-LM.}
\label{tab:resnet}
\begin{tabular}{|c|c|c|c|c|c|}
\hline
ResNet layers & 2     & 4     & 6     & 8     & 10    \\ \hline
samples pass CN+EB check(\%)  & 42.19 & 42.92 & 43.84 & 42.76 & 42.23 \\ \hline
\end{tabular}
\end{table}

Word frequency constitutes a pivotal parameter in the Diffusion-BERT model. When introducing noise into a sequence, this parameter delineates the extent to which noise relies on the frequency of each token present in the training set. In our study, we conducted a comparative analysis of samples generated by the Diffusion-BERT models with word frequency values ranging from 0.1 to 1. The corresponding percentages of samples that successfully passed the CN+EB check are presented in Table \ref{tab:word_freq}. As we progressively augmented the value of word frequency, we observed an initial increase followed by a subsequent decline in the percentage of generated samples meeting the criteria for chemical stability. Remarkably, we identified that the model achieved its peak performance, reaching 83.00\%, when the word frequency was set to 0.3.

\begin{table}[ht]
\centering
\caption{Ablation studies for word frequency used in Diffusion-BERT.}
\label{tab:word_freq}
\begin{tabular}{|c|c|c|c|c|c|c|c|c|c|c|}
\hline
word frequency:   & 0.1  & 0.2  & 0.3  & 0.4  & 0.5  & 0.6  & 0.7  & 0.8  & 0.9  & 1    \\ \hline
\makecell{samples pass \\ CN+EB check(\%)}    & 77.74 & 75.60 & 83.00 & 82.03 & 81.88 & 80.98 & 76.72 & 72.88 & 78.26 & 75.62 \\ \hline
\end{tabular}
\end{table}

\section{Results}
\subsection{Dataset}

Our diffusion-based generation model is trained using the compositions of materials obtained from the Materials Project (MP) database \cite{MP}. Since this database assigns a unique ID to each structure, rather than compositions, and our generation model specifically focuses on the compositions, we must perform additional data filtering to remove duplicate compositions. We exclude lanthanides, actinides, and noble gas elements from consideration, as they may not be suitable for template-based element substitution in crystal structure prediction. By eliminating duplicate compositions and compositions containing these elements, we curate a dataset comprising 93,307 distinct formulas. Furthermore, as we are not supposed to get too complicated compositions from our generation models, we select a subset of 59,330 unique formulas, each containing fewer than 50 atoms, as our final training set.

\subsection{Composition generation performance}

Table \ref{table:formula_results} shows the composition generation performances of our diffusion models. Out of a total of 20,000 sequences generated by Diffusion-LM and Diffusion-BERT, 6,351 and 9,814 clean formulas with no special tokens within the token sentences were obtained, respectively. After removing duplicate formulas, the Diffusion-LM and Diffusion-BERT generated 5,378 and 9,814 unique formulas, with 5,067 and 9,768 of these unique formulas not appearing in the MP database, respectively. The recovery rates of formulas generated by these two methods are 2.78\% and 0.14\%, respectively, indicating our models' capability to discover experimentally verified materials. Furthermore, 4,384 and 8,300 generated formulas successfully passed the charge-neutrality and electronegativity balance checks,  while 1,976 and 1,421 formulas respectively met the final oxidation states check by BERTOS. The above performance illustrates that Diffusion-BERT excels in generating a more diverse range of formulas, whereas Diffusion-LM tends to produce formulas with greater chemical validity.

\begin{table*}[ht]
\centering
\caption{Generation results of Diffusion-LM and Diffusion-BERT}
\label{table:formula_results}
\begin{tabular}{|c|c|c|c|c|c|c|c|c|}
\hline
Model  &  Generated  &  Formula  &  Unique  &  Recover rate  &  Novelty      & CN   & EB   & OS \\ \hline
Diffusion-LM   & 10,000    & 63.51\%    & 53.78\%      & 2.78\%     & 50.67\%    & 45.94\%  &  43.84\%  & 19.76\% \\ \hline
Diffusion-BERT & 10,000    & 98.14\%    & 98.14\%      & 0.14\%     & 97.98\%    & 84.31\%  &  83.00\%  & 14.21\% \\ \hline
\end{tabular}
\end{table*}

Table \ref{table:methods} shows the performance of our diffusion language model-based composition generator in comparison to other GAN-based and large language model-based generators. Only 73.4\% of samples generated by MATGAN, a GAN-based composition generator trained with the MP database, were successful in passing both CN and EB checks. 
Furthermore, our Diffusion-BERT exhibits superior performance to four out of six transformer-based composition generators in terms of the percentage of samples that clear both checks. Only MT-GPTJ and MT-GPTNeo outperform our Diffusion-BERT. These findings underscore the competitiveness of the diffusion language model-based composition generator.

\begin{table*}[ht]
\centering
\caption{Comparison of generator performances with GAN \cite{dan2020generative} and Crystal Transformer \cite{wei2022crystal}}
\label{table:methods}
\begin{tabular}{|c|c|c|c|c|c|c|c|c|}
\hline
Model  &  CN   & CN + EB \\ \hline
MatGAN \cite{dan2020generative}        & 84.8\%   &  73.4\%   \\ \hline
MT-GPT \cite{wei2022crystal}           & 92.24\%  &  50.61\%  \\ \hline
MT-GPT2 \cite{wei2022crystal}          & 92.96\%  &  79.79\%  \\ \hline
MT-GPTNeo \cite{wei2022crystal}        & 93.84\%  &  84.37\%  \\ \hline
MT-GPTJ \cite{wei2022crystal}          & 96.98\%  &  90.33\%  \\ \hline
MT-BART \cite{wei2022crystal}          & 81.10\%  &  62.83\%  \\ \hline
MT-RoBERTa \cite{wei2022crystal}       & 71.16\%  &  61.00\%  \\ \hline
Diffusion-LM                           & 45.94\%  &  43.84\%  \\ \hline
Diffusion-BERT                         & 84.31\%  &  83.00\%  \\ \hline

\end{tabular}
\end{table*}

\subsection{Structure generation results}

As ternary and quaternary materials consist of a majority of functional materials, our structure generation process focuses on these materials. For ternary and quaternary composition samples generated by our Diffusion-LM and Diffusion-BERT models, we first check the ElMD \cite{ElMD} score between a new formula and all formulas existing in the MP database and then choose the one with the smallest ElMD distance as the template for CSPML \cite{CSPML} algorithm to generate corresponding structures. Notably, approximately 60\% of the generated samples yield corresponding structures that cannot be predicted using the CSPML method, mainly due to their elemental ratios not being present in the MP database. These outlier formulas also indicate that the outputs of our diffusion models have high diversity. Following this, we employ M3GNet \cite{M3GNET} to further relax these structures and calculate the formation energies of the relaxed structures.

\begin{figure*}[ht] 
    \centering
    \begin{minipage}[c]{0.4\textwidth}
        \centering
        \includegraphics[width=\textwidth]{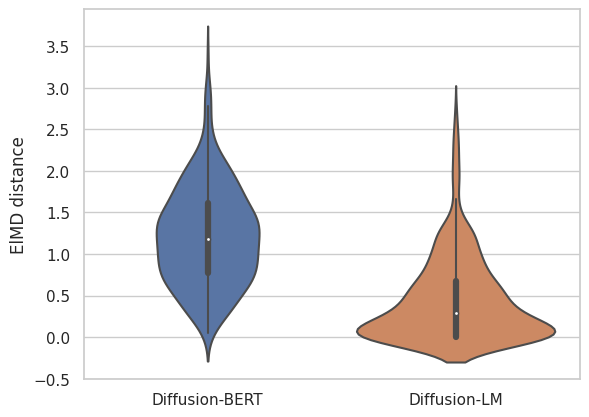}
        \subcaption{}
    \end{minipage}
    \begin{minipage}[c]{0.4\textwidth}
        \centering
        \includegraphics[width=\textwidth]{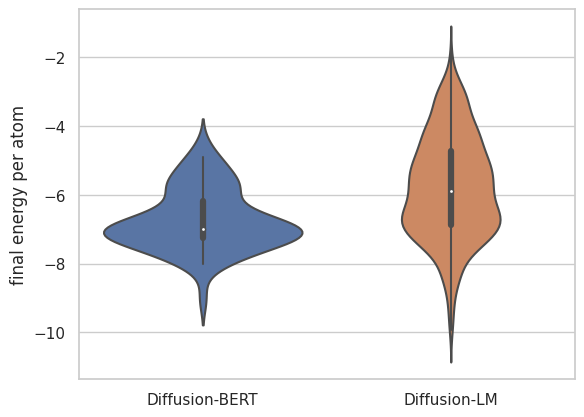}
        \subcaption{}
    \end{minipage}\\
   
    \caption{Performance of generation results of Diffusion-BERT and Diffusion-LM. (a) ElMD distance distribution of generated formulas. (b)Formation energy per atom distribution of generated structures relaxed by M3GNet.}
    \label{fig: distribution}
\end{figure*}

The ElMD distributions of Diffusion-LM and Diffusion-BERT are visualized in Fig \ref{fig: distribution} (a). It is evident that Diffusion-LM-generated formulas exhibit considerably smaller ElMD distances compared to those generated by Diffusion-BERT, implying that Diffusion-LM's outputs are more analogous to known formulas present in the MP database. Consequently, when we apply a template-based structure prediction method to these generated formulas, structures derived from Diffusion-LM's formulas are more likely to exhibit stability in comparison to those from Diffusion-BERT. Meanwhile, the formulas generated by Diffusion-BERT showcase a higher level of diversity relative to Diffusion-LM. Subsequent to structure relaxation, we utilize the formation energy per atom as a parameter to evaluate the stability of these newly generated structures. The formation energy per atom distributions of Diffusion-LM and Diffusion-BERT are depicted in Figure \ref{fig: distribution} (b). Structures generated based on Diffusion-BERT's formulas display lower formation energy values compared to those generated based on Diffusion-LM's formulas, indicating a potential for identifying more thermodynamically stable materials among the candidates produced by Diffusion-BERT.

\subsection{New structures predicted by our algorithms}

\begin{figure*}[ht] 
    \centering
    \begin{minipage}[c]{0.3\textwidth}
        \centering
        \includegraphics[width=\textwidth]{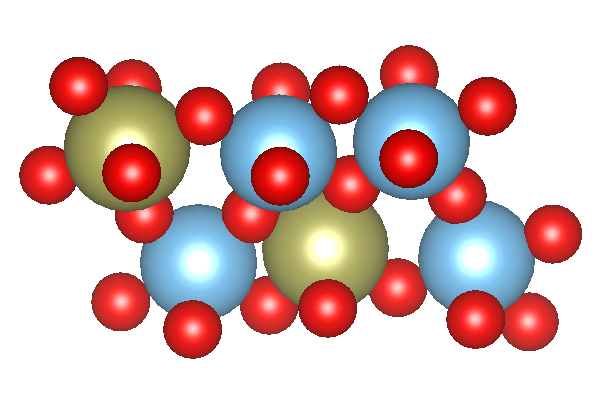}
        \subcaption{}
    \end{minipage}
    \begin{minipage}[c]{0.3\textwidth}
        \centering
        \includegraphics[width=\textwidth]{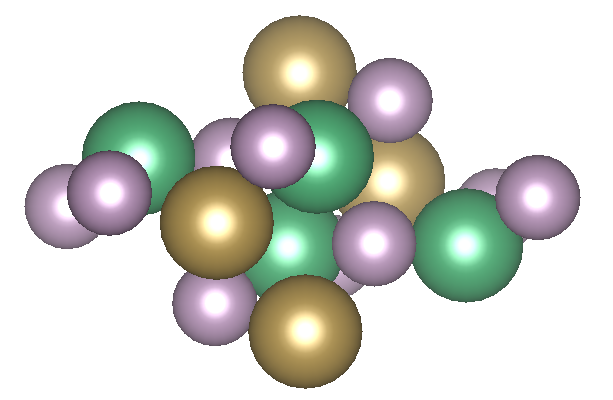}
        \subcaption{}
    \end{minipage}\\
    \begin{minipage}[c]{0.3\textwidth}
        \centering
        \includegraphics[width=\textwidth]{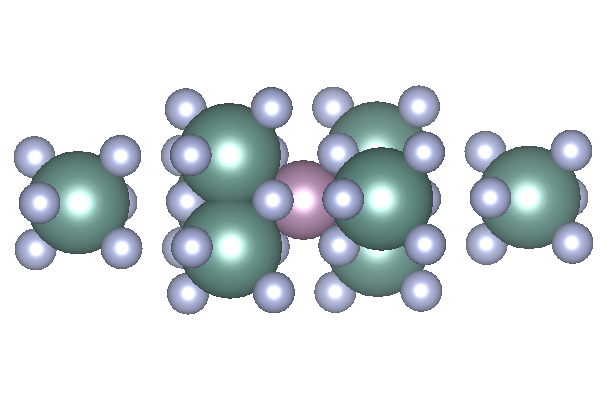}
        \subcaption{}
    \end{minipage}
    \begin{minipage}[c]{0.3\textwidth}
        \centering
        \includegraphics[width=\textwidth]{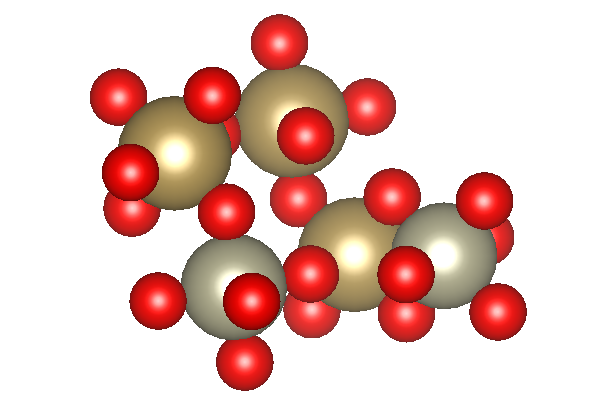}
        \subcaption{}
    \end{minipage}\\
    \caption{DFT-relaxed new structures. (a) Ti$_2$HfO$_5$. (b) TaNbP. (c) YMoN$_2$. (d) TaReO$_4$.}
    \label{fig: structures}
\end{figure*}

\begin{figure}[ht] 
    \centering
    \begin{minipage}[c]{0.3\textwidth}
        \centering
        \includegraphics[width=\textwidth]{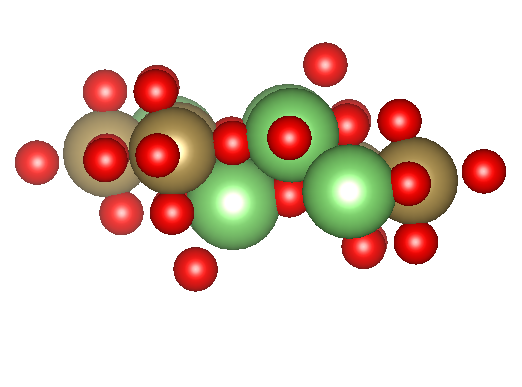}
        \subcaption{Structure candidate for Li$_2$Ta$_3$O$_7$}
    \end{minipage}
    \begin{minipage}[c]{0.3\textwidth}
        \centering
        \includegraphics[width=\textwidth]{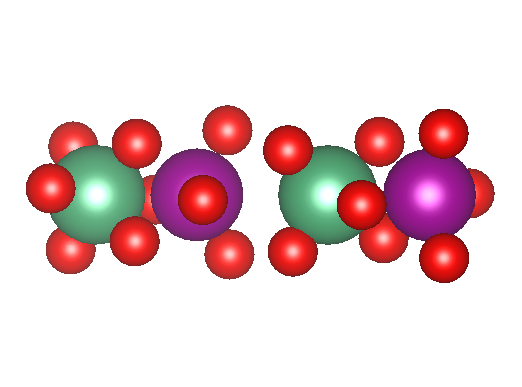}
        \subcaption{Structure candidate for MnNbO$_3$}
    \end{minipage}
    \begin{minipage}[c]{0.3\textwidth}
        \centering
        \includegraphics[width=\textwidth]{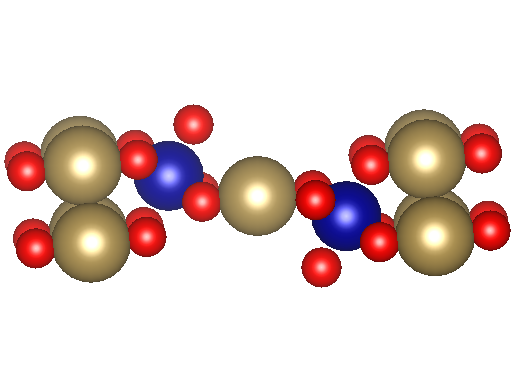}
        \subcaption{Structure candidate for TaCrO$_2$}
    \end{minipage}\\
    \begin{minipage}[c]{0.3\textwidth}
        \centering
        \includegraphics[width=\textwidth]{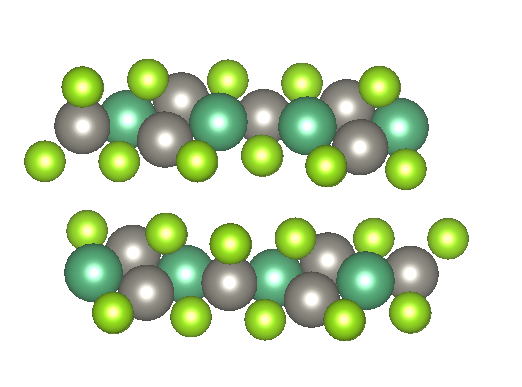}
        \subcaption{Structure candidate for Nb$_2$W$_3$Se$_5$}
    \end{minipage}
    \begin{minipage}[c]{0.3\textwidth}
        \centering
        \includegraphics[width=\textwidth]{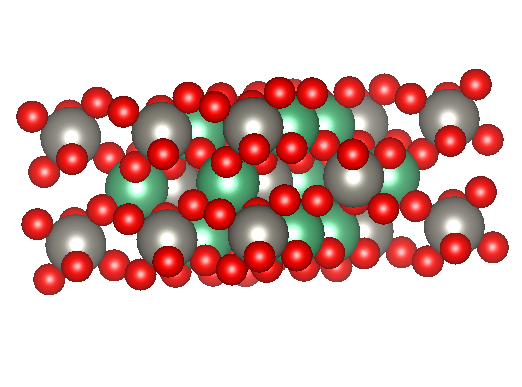}
        \subcaption{Structure candidate for Nb$_2$WO$_6$}
    \end{minipage}
    \begin{minipage}[c]{0.3\textwidth}
        \centering
        \includegraphics[width=\textwidth]{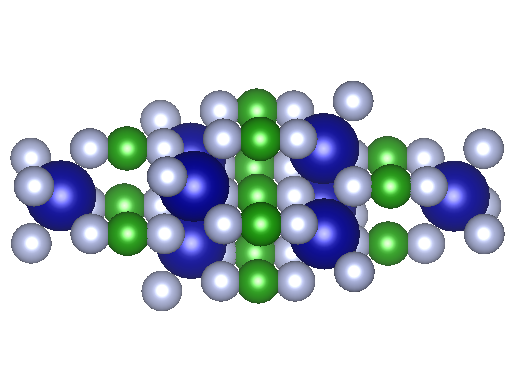}
        \subcaption{Structure candidate for CrBN$_2$}
    \end{minipage}\\
     \begin{minipage}[c]{0.3\textwidth}
        \centering
        \includegraphics[width=\textwidth]{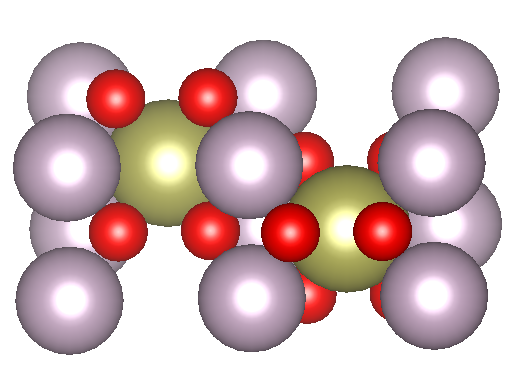}
        \subcaption{Structure candidate for HfTcO$_2$}
    \end{minipage}
    \begin{minipage}[c]{0.3\textwidth}
        \centering
        \includegraphics[width=\textwidth]{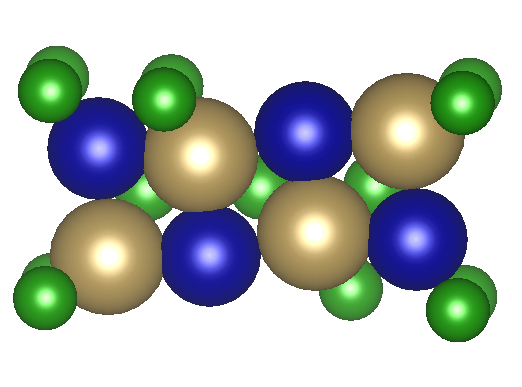}
        \subcaption{Structure candidate for Ta$_2$Cr$_2$B$_3$}
    \end{minipage}
    \begin{minipage}[c]{0.3\textwidth}
        \centering
        \includegraphics[width=\textwidth]{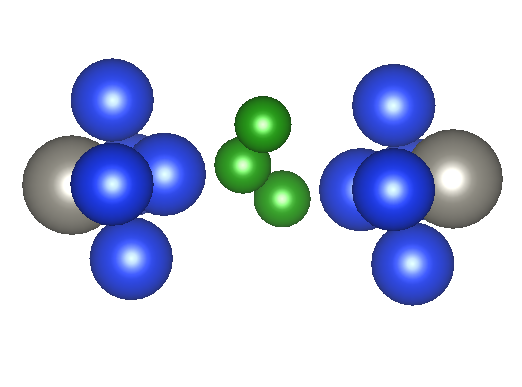}
        \subcaption{Structure candidate for SiBW}
    \end{minipage}\\
    \caption{New material candidates generated from our pipeline.}
    \label{fig:newstructure2}
\end{figure}

For new formulas generated through Diffusion-LM and Diffusion-BERT algorithms, we first apply CSPML to obtain initial candidate structures. These structures are then relaxed by M3GNet, followed by validation using DFT. Among the 3,392 formulas that have successfully passed the charge-neutral, electronegativity balance, and oxidation state checks, we identified 2,214 structures with formation energies predicted by M3GNet below zero. We have included the top 1000 generated formulas with the lowest formation energy in Supplementary Table S1. Additionally, we employ DFT calculations to relax the top ten structures with the lowest formation energy, revealing that six out of ten exhibited formation energies less than zero. Furthermore, four of these six structures displayed e-above-hull energy of less than 0.3 eV (the threshold used in \cite{lyngby2022data}). The e-above-hull energy and formation energy for these six new materials are presented in Table \ref{tab:ehull}. Figure \ref{fig: structures} showcases four DFT-relaxed new structures discovered by our model with e-above-hull energy values below 0.3 eV. Additionally, Figure \ref{fig:newstructure2} illustrates an additional nine hypothetical materials generated by our pipeline with M3GNET-predicted formation energies less than 0.

\begin{table}[ht!]
\centering
\caption{E-above-hull energy and formation energy of selected new materials.}
\label{tab:ehull}
\begin{tabular}{|c|c|c|}
\hline
formulas & E-above-hull energy (eV) & Formation energy (eV/atom)\\ \hline
Ti$_2$HfO$_5$ & 0.018 & -3.411\\ \hline
TaNbP & 0.093 & -0.702\\ \hline
YMoN$_2$ & 0.123 & -0.930 \\ \hline
TaReO$_4$ & 0.144 & -2.422\\ \hline
HfTiO$_2$ & 0.503 & -2.434\\ \hline
HfMnO$_2$ & 0.577 & -2.248 \\ \hline
\end{tabular}
\end{table}

\section{Conclusion}

To explore novel inorganic materials, the initial step is to generate new chemically stable compositions. Subsequently, thermodynamically stable structures are produced based on these compositions. Here we introduce a diffusion-based pipeline for generating novel materials by amalgamating diffusion language model-based composition generators, a template-based crystal structure predictor, and a graph neural network potential-based structure relaxation algorithm. Comparison with previous GAN-based and transformer-based composition generators reveals the superior performance of our diffusion-based composition generator. It achieves results that outperform the GAN-based generator and demonstrates competitive performance with transformer-based composition generators. Through DFT validation, we have identified six newly discovered material structures, all exhibiting formation energies below zero. Notably, Ti$_2$HfO$_5$, TaNbP, YMoN$_2$, and TaReO$_4$ exhibit e-above-hull energies below 0.3 eV. These findings underscore the potential of our diffusion-based approach for discovering novel structures.

\section{Data and Code Availability}
The dataset is downloaded from the Materials Project database using its API. The source code is available from the corresponding author upon reasonable request.

\section{Conflict of interest}
There are no conflicts to declare.

\section{Contribution}
Conceptualization, J.H.; methodology, R.D., E.S., J.H.; software, R.D., J.H.; resources, J.H.; writing--original draft preparation, R.D., J.H., N.F., E.S.; writing--review and editing, J.H; visualization, R.D.; supervision, J.H.; funding acquisition, J.H.

\section*{Acknowledgement}
The research reported in this work was supported in part by the National Science Foundation under grant 2110033. The views, perspectives, and content do not necessarily represent the official views of the NSF.

\bibliographystyle{unsrt}  
\bibliography{references}

\end{document}